# Unconventional Hall effect and its variation with Co-doping in van der Waals Fe$_3$GeTe$_2$

Rajeswari Roy Chowdhury[1]*, Samik DuttaGupta[2-4]*, Chandan Patra[1], Oleg A. Tretiakov[5], Sudarshan Sharma[1], Shunsuke Fukami[2-4,6,7], Hideo Ohno[2-4,6,7] and Ravi Prakash Singh[1]*

[1] *Department of Physics, Indian Institute of Science Education and Research Bhopal, Bhopal Bypass Road, Bhauri, Madhya Pradesh 462-066, India*

[2] *Center for Science and Innovation in Spintronics, Tohoku University, 2-1-1 Katahira, Aoba-ku, Sendai 980-8577, Japan*

[3] *Center for Spintronics Research Network, Tohoku University, 2-1-1 Katahira, Aoba-ku, Sendai 980-8577, Japan*

[4] *Laboratory for Nanoelectronics and Spintronics, Research Institute of Electrical Communication, Tohoku University, 2-1-1 Katahira, Aoba-ku, Sendai 980-8577, Japan*

[5] *School of Physics, The University of New South Wales, Sydney 2052, Australia*

[6] *Center for Innovative Integrated Electronic Systems, Tohoku University, 468-1 Aramaki Aza Aoba, Aoba-ku, Sendai 980-0845 Japan*

[7] *WPI Advanced Institute for Materials Research, Tohoku University, 2-1-1 Katahira, Aoba-ku, Sendai 980-8577, Japan*

*Corresponding author: rajeswari@iiserb.ac.in, sdg@riec.tohoku.ac.jp, rpsingh@iiserb.ac.in

**Two-dimensional (2D) van der Waals (vdW) magnetic materials have attracted a lot of attention owing to the stabilization of long-range magnetic order down to atomic dimensions, and the prospect of novel spintronic devices with unique functionalities. The clarification of the magnetoresistive properties and its correlation to the underlying magnetic configurations is essential for 2D vdW-based spintronic devices. Here, the effect of Co-doping on the magnetic and magnetotransport properties of $Fe_3GeTe_2$ have been investigated. Magnetotransport measurements reveal an unusual Hall effect behavior whose strength was considerably modified by Co-doping and attributed to arise from the underlying complicated spin textures. The present results provide a clue to tailoring of the underlying interactions necessary for the realization of a variety of unconventional spin textures for 2D vdW FM-based spintronics.**

**Introduction**

Two-dimensional (2D) van der Waals (vdW) materials have recently drawn considerable attention owing to their prospect for 2D magnetism, spintronic and magneto-optical applications[1-4]. The stabilization of a long-range ferromagnetic and/or antiferromagnetic order down to the atomic limit and its seamless flexibility with various elements and structures provide an exciting opportunity for exploring new physical properties and realization of novel electronic devices down to the 2D limit[2-4]. Among the family of layered vdW ferromagnets (FMs), quasi-2D metallic $Fe_3GeTe_2$ is a promising candidate owing to its relatively high Curie temperature (~ 220 K)[5-10], large anomalous Hall effect (AHE)[7-11], and significant uniaxial magnetic anisotropy[12], down to the atomic limit. Figure 1a shows the crystal structure of $Fe_3GeTe_2$. It possesses a hexagonal structure (see Fig. 1b) (space group- $P6_3/mmc$)[5,6,13], where a $Fe_3Ge$ substructure containing inequivalent Fe atoms is sandwiched between two vdW-bonded Te layers. Previous investigations using single-crystals and nanometer-sized flakes have demonstrated that the competition between magnetic exchange and dipole-dipole interaction, relativistic band structure effects, and magnetic frustration originating from inequivalent Fe atoms manifest in topological nodal line-driven large anomalous Hall angle[11], and formation of various domain structures including stripe domains[14] and hexagonal lattice of skyrmion bubbles[15,16]. Besides, an unconventional Hall effect anomaly has been reported[9,17,18], whose origin is heavily debated and attributed to a multitude of factors, including the formation of in-plane skyrmions[17,19], non co-planar spin configurations[9], or multi to single-domain transformations under applied magnetic field[18]. Thus, an understanding of the factors determining the origin of the unusual magnetotransport features and its possible correlation with an underlying domain or topological spin textures is essential for the subsequent development of 2D vdW FM-based spintronic devices. One of the possible ways to achieve this objective pertains to the chemical substitution of the magnetic (Fe) site by a different element, which is expected to modify the underlying interactions and enable understanding of the factors contributing to the unconventional magnetotransport behavior.

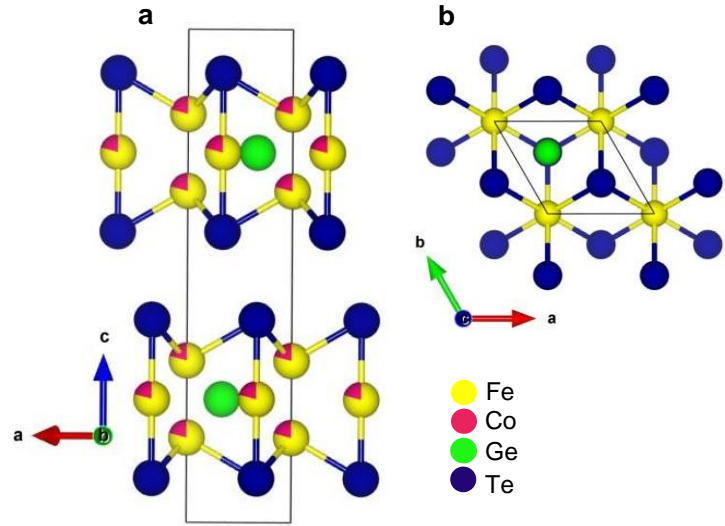

**Fig. 1.** (a) Crystal structure of $Fe_3GeTe_2$. The Co atoms (red circles) are distributed randomly at the Fe site (yellow circles). (b) Crystal structure as viewed from *ab* plane with the *c*-axis along out-of-the plane of the paper.

Some previous studies have shown that the substitution of Fe site by Co results in substantial changes of the magnetic properties (*e.g.*, ferromagnetic transition temperature)[20], modification of magnetic anisotropy strength[21], and the emergence of a hard-magnetic phase with strong domain-wall pinning effects[20]. However, the effect of Co-doping on the magnetoresistive properties and the associated variations in the underlying magnetic domain and/or topological spin textures have not been clarified yet.

Here, we show the impact of Co-doping on the unconventional Hall effect in the uniaxial vdW FM $Fe_3GeTe_2$. Magnetotransport measurements demonstrate a substantial modification of an unconventional cusp-like behavior in Hall resistivity with Co-doping, tentatively attributed to underlying domain structures from complementary magneto-optical investigations. The obtained results demonstrate a possible route towards the tuning of the properties within the family of vdW FMs, prospective for topological magnetism, and for the realization of a variety of non-collinear spin textures at reduced dimensions.

**Results**

**Single-crystal growth and characterization.** Single crystalline $(Co_xFe_{1-x})_3GeTe_2$ ($Co_x$FGT, hereafter) samples and a reference sample of $Fe_3GeTe_2$ (FGT, hereafter) were grown by chemical vapor transport (CVT) method. The various doping levels of 0, 0.05, 0.45, and 0.55 utilized in this study refer to the Co concentrations ($x$) determined using the atomic percentage ratios obtained from scanning electron microscope (SEM) energy-dispersive X-ray (EDX) spectroscopy measurements. X-ray diffraction (XRD) was carried out at room temperature with Cu-$K_\alpha$ radiation. Figure 2a shows the experimental results of out-of-plane XRD for undoped, slightly doped ($Co_{0.05}$FGT), and significantly doped ($Co_{0.45}$FGT) single-crystals. The observed Bragg peaks can be indexed with (00$l$) peaks, indicating that the sample surface corresponds to the *ab* plane with the *c*-axis along the out-of-plane direction. Besides, any new peak corresponding to unreacted elements or peak shifts due to unintentional formation of other Co-Fe variants was not observed, indicating a homogenous mixing of Co and Fe. Laue diffraction pattern of the FGT crystal (see Fig. 2b) confirms the formation of high-quality single crystals. The obtained $c$-axis lattice parameter from XRD shows a monotonic decrease with increasing $x$ (see Fig. 2c), consistent with earlier reports[20]. The EDX spectra of the samples, as shown in Fig. 2d-f confirms Co, Fe, Ge, and Te elements in the $Co_xFGT$ and FGT samples, respectively. The magnetic properties were characterized using a superconducting quantum interference device within the temperature range 5-300 K. Polar magneto-optical Kerr effect (MOKE) investigations were carried out on freshly cleaved *ex-situ* samples. Electrical and magnetotransport measurements were performed by a physical property measurement system using a conventional four-probe technique. To check the reproducibility of the observed results, similar experiments were repeated in separate single crystals of different sizes. Longitudinal ($\rho_{XX}$) and transverse ($\rho_{XY}$) resistivities were obtained as a function of both temperature and applied magnetic field ($H$).

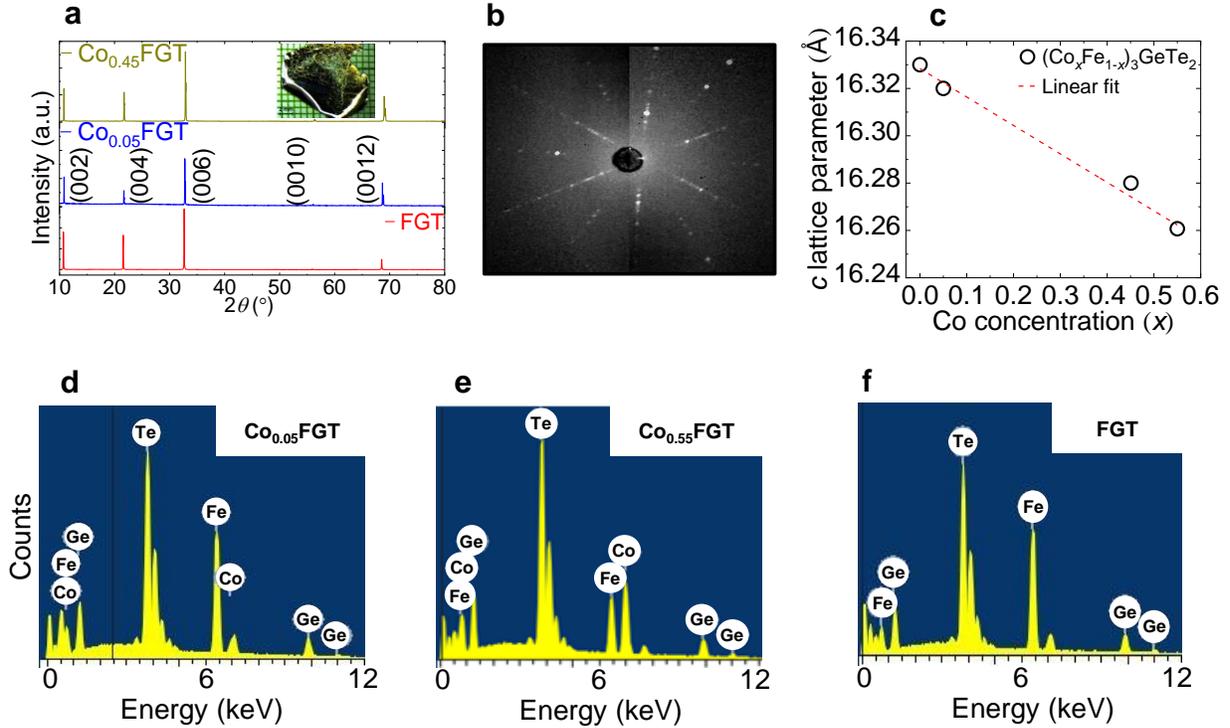

**Fig. 2.** (a) Out-of-plane X-ray diffraction pattern for $Co_{0.05}FGT$ (blue curve), $Co_{0.45}FGT$ (dark yellow curve), and FGT (red curve) single-crystalline samples at room temperature. The inset in (a) shows the optical micrograph of a cm sized FGT single-crystal utilized in this study. (b) Laue diffraction pattern of FGT single crystal. (c) Variation of $c$-axis lattice parameter versus Co-concentration ($x$) determined from energy dispersive X-ray (EDX) spectroscopy. (d)-(f) EDX spectra for $Co_{0.05}FGT$, $Co_{0.55}FGT$, and FGT, respectively.

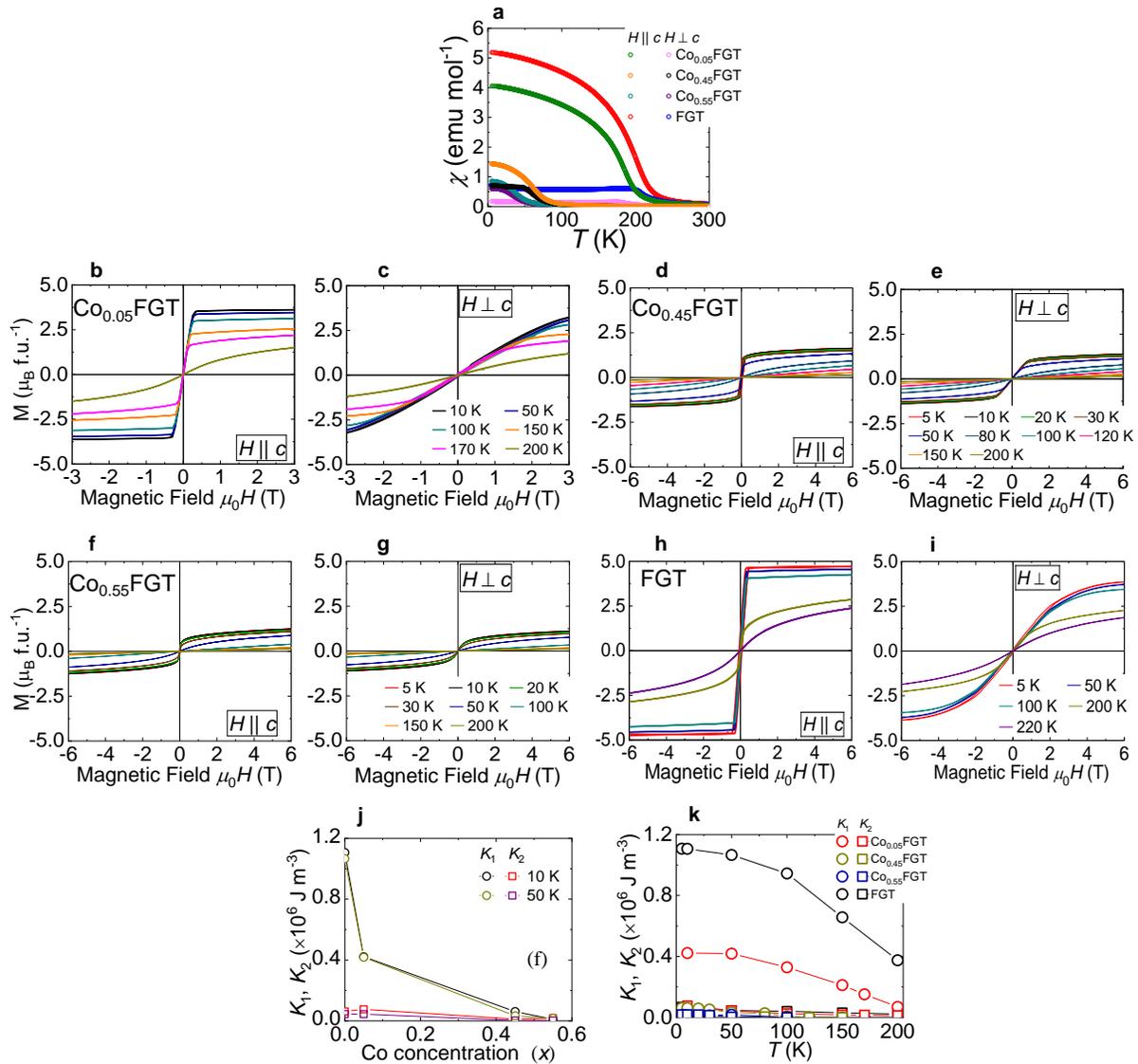

**Fig. 3.** (a) Magnetic susceptibility ($\chi$) versus temperature ($T$) under magnetic field $\mu_0 H = 0.5$T applied parallel to *c*-axis and perpendicular to *c*-axis for the doped (Co$_x$FGT) and undoped (FGT) single-crystalline

samples. (b), (c) Field ($H$) dependence of magnetization ($M$) for $Co_{0.05}$FGT single-crystal at various temperatures with $H \parallel c$-axis and $H \perp c$-axis, respectively. (d), (e) $H$ dependence of magnetization $M$ for $Co_{0.45}$FGT single-crystal at various temperatures with $H \parallel c$-axis and $H \perp c$-axis, respectively. (f), (g) $H$ dependence of magnetization $M$ for $Co_{0.55}$FGT single-crystal at various temperatures with $H \parallel c$-axis and $H \perp c$-axis, respectively. (h), (i) $M$ versus $H$ for the reference FGT single-crystalline sample at indicated temperatures for $H \parallel c$-axis and $H \perp c$-axis, respectively. (j) Co-concentration ($x$) dependence of first-order ($K_1$) and second-order ($K_2$) anisotropy constants at $T$ = 10 and 50 K. (k) $T$ dependence of first-order ($K_1$) and second-order ($K_2$) anisotropy constants for $Co_x$FGT and reference FGT single-crystalline samples.

**Magnetic Properties of $Co_x$FGT and FGT.** To investigate the impact of doping on the magnetic properties, we measured the temperature ($T$) dependence of magnetization and magnetic hysteresis of the single crystals. Figure 3a shows the experimental results of field-cooled (FC) magnetic susceptibility ($\chi$) vs. $T$, for $Co_x$FGT and FGT, under an applied magnetic field $\mu_0 H$ = 0.5 T, parallel (out-of-plane) and perpendicular (in-plane) to $c$-axis ($\mu_0$ is the permeability in vacuum). For FGT, upturns in $\chi$ were observed around $T \approx$ 220 K for $H \parallel c$-axis and $T \approx$ 200 K for $H \perp c$-axis, respectively, demonstrating the onset of ferromagnetic ordering. The introduction of Co results in a gradual decrease of ferromagnetic transition temperature ($T_C$) to ~ 200 K ($H \parallel c$-axis) and 170 K ($H \perp c$-axis) for $Co_{0.05}$FGT, ~ 100 K and 50 K for $Co_{0.45}$FGT, and ~ 50 K (both for $H \parallel c$ and $H \perp c$-axis ) for $Co_{0.55}$FGT. Besides, an increase of $x$ also results in a suppressed bifurcation of $\chi$ along $H \parallel c$ and $H \perp c$-axes, representing a significant reduction of the anisotropic magnetic character. Figure 3b-i shows the magnetic hysteresis curves along the out-of-plane ($H \parallel c$) and in-plane ($H \perp c$) directions for $Co_x$FGT and reference FGT. The $M$-$H$ curves for $H \parallel c$-axis show the magnetic easy-axis along the $c$-axis and a lowering of the spontaneous magnetic moment with increasing $x$. The observed suppression of $T_C$ and spontaneous magnetic moment is consistent with the previous experimental results on transition metal doped FGT single-crystals[20,22]. To clarify the influence of doping on magnetic anisotropy, *M-H* curves for $H \perp c$-axis were utilized to determine first-order ($K_1$) and second-order ($K_2$) anisotropy constants by Sucksmith-Thompson method[8,23]. Figure 3j shows the Co concentration ($x$) dependence of $K_1$ and $K_2$ at 10 and 50 K ($T < T_C$ for all the samples). An increase of $x$ results in a considerable weakening of $K_1$ by more than an order of magnitude while still retaining a positive value, indicating the uniaxial character of the magnetic anisotropy. Furthermore, for both Co$_x$FGT and FGT, the temperature dependence of anisotropy constants shows a monotonic decrease of $K_1$ with a rise of temperature. On the other hand, no significant effect of doping or temperature on $K_2$ was observed (see Fig. 3k). The observed trend of variation of $K_1$ can be qualitatively explained to arise from sharp changes in the magnetic anisotropy energy associated with spin-orbit interaction induced changes of electronic structure with reduced Fe concentration[21] and/or Co doping, while that for $K_2$ indicates the presence of additional factors. Assuming $K_1$ is equivalent to uniaxial anisotropy constant ($K_U$), we determine the quality factor Q = $2K_U/\mu_0 M_S^2$ (where $M_S$ is saturation magnetization). We obtain Q = 7.58, 7.21, 4.29 and 11.98 for Co$_{0.05}$FGT, Co$_{0.45}$FGT, Co$_{0.55}$FGT, and FGT, respectively (at 10 K). The obtained values are lower than some previous reports[8,9,18], indicating a relatively soft ferromagnetic nature with Co doping. The results from the magnetization measurements will be later used for the extraction of the unconventional Hall contribution from $\rho_{XY}$ vs. magnetic field curves.

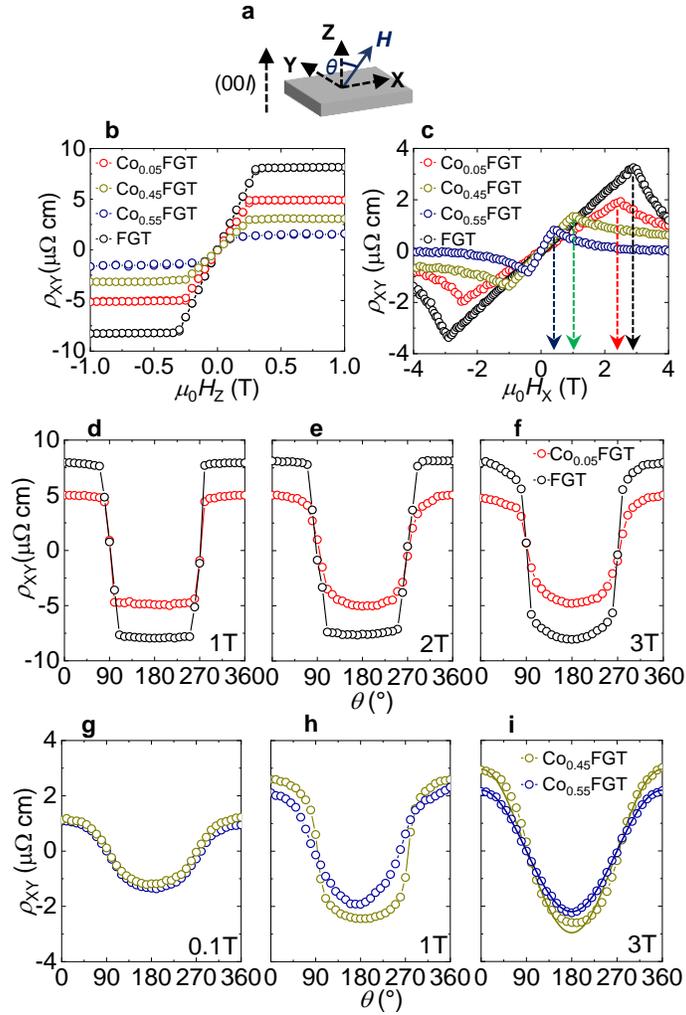

**Fig. 4.** (a) Schematic representation of the measurement configuration utilized in this work. The angle ($\theta$) is defined as the angle subtended by the applied magnetic field ($H$) with respect to the crystallographic $c$-axis (out-of-plane) of the single-crystalline samples. Applied current ($I$) for measurement is along X-direction. (b) Hall resistivity ($\rho_{XY}$) versus $H$ ($||$ $c$-axis) for $Co_{0.05}FGT$, FGT at 100K and $Co_{0.45}FGT$, $Co_{0.55}FGT$ at 10 K. (c) $\rho_{XY}$ versus $H$ ($\perp c$ $||$ $I$) for $Co_{0.05}FGT$, FGT at 100K and $Co_{0.45}FGT$, $Co_{0.55}FGT$ at 10 K. The arrows in the figure denotes the position of maximum in $\rho_{XY}$ for the single crystals investigated in this study. (d), (e), (f) $\theta$ dependence of $\rho_{XY}$ for $Co_{0.05}FGT$ and FGT under applied $\mu_0 H$ = 1, 2 and 3 T at 100 K. (g), (h), (i) Experimental results for similar measurements on $Co_{0.45}FGT$ and $Co_{0.55}FGT$ under applied $\mu_0 H$ = 0.1, 1 and 3 T at 10 K. Solid lines in (i) denotes the fitting of the experimental data with the harmonic cosine *θ* dependence.

**Unconventional Hall effect and its modification with Co-doping in FGT.** Until now, most of the studies concerning the formation of topological spin textures have focused on the utilization of relativistic chiral Dzyaloshinskii-Moriya interaction (DMI)[24,25], either originating from crystallographic lattice lacking inversion symmetry in single-layered FMs[26-33] or due to broken space inversion symmetry in ferromagnetic multilayers[34-37]. Owing to the non-trivial nature of the spin structures, a traversing electron experiences a continuous precession of its spin magnetic moment, equivalent to a virtual (or emergent) magnetic field. Conventionally, the effect of this emergent field manifests as an additional contribution to Hall resistivity and serves as an electrical signature of the existence of topological spin textures, commonly referred to as topological Hall effect (THE)[27,38-42]. On the other hand, the possibility of utilization of centrosymmetric FMs as topological spin texture hosting materials was completely neglected owing to symmetry requirements, which dictates the absence of DMI in these systems. Recent experimental results have demonstrated the formation of topological skyrmion lattice and/or chiral structures in single-crystalline bulk FGT[15,43], exfoliated nanoflakes[16,44], and formation of Néel-type skyrmions in FGT-based heterostructures[44-46]. However, the experimental results from the magnetoresistive measurements[9,17,18] cannot be solely interpreted to arise from submicron skyrmion bubble-like textures, demonstrated by magnetic imaging techniques[15]. In a practical scenario, doping of the Fe site by Co might modify the underlying magnetic microstructure and stabilize or destabilize topological spin textures by additional symmetry-breaking interactions[44], enabling a resolution of the previous experimental results and tuning of topological spin structures in this material system. Figure 4a shows the schematic diagram of the measurement geometry utilized in this work. An applied dc *I* (⊥ *c*-axis) of magnitude 10 mA was passed through the single-crystals along the *x*-direction (*i.e.*, along crystal *ab* plane). The resulting change in $\rho_{XY}$ under simultaneous application of external $H$ (along $z$ or $x$ directions) and $I$ was obtained by measuring voltage drop along the $y$-direction. For applied $H_Z$ ($\parallel c$-axis), we have observed a sizable Hall resistance, which can be attributed to the anomalous Hall effect (AHE)[9,10,18], possibly originating from topological nodal lines in the band structure with the magnetization pointing along the easy axis[11] (see Fig. 4b). The magnitude of AHE decreases with the increase of $x$, owing to the reduction of spontaneous magnetization with increasing Co-concentration (see Fig. 3b,d,f,h), while no characteristic features of THE were observed in $\rho_{XY}$ vs. $H_Z$. Furthermore, $\rho_{XY}$ measurements also manifest in low magnitudes of the coercive field and remanence to saturation ratios, confirming the soft ferromagnetic nature. On the other hand, for the applied field along $x$-direction ($H_X \parallel I \perp c$-axis), we have observed a reduction of $\rho_{XY}$ magnitude along with the emergence of a prominent cusp-like feature (see Fig. 4c). The position of the cusp-like feature is independent of the in-plane field direction ($H_X$ or $H_Y$) (see supplementary figure S1) and shifts towards lower field values with increasing $x$. To further examine the origin of the observed features, we have also measured $\rho_{XY}$ as a function of the angle $\theta$ for various applied field magnitudes ($\theta$ is defined as the angle subtended by $H$ with respect to the out-of-plane $z$-direction) (see Fig. 4d-i). For applied $\mu_0 H_X \leq 2$ T, $\rho_{XY}$ sharply deviates from the usual cosine $\theta$ relation expected from conventional Hall effect[47], hinting towards a different origin of the observed magnetotransport behavior (see Fig. 4d,e,g,h). An increase of $H$ ($\mu_0 H_X = 3$ T) resulted in a gradual transition towards the conventional harmonic nature for $Co_{0.05}FGT$, $Co_{0.45}FGT$, and FGT, possibly due to a transition towards a single-domain behavior. A further increase of $H$ is expected to result in a cosine $\theta$ behavior, where the magnetization follows the rotating $H$. This situation is observed for $Co_{0.55}FGT$ (see Fig. 4i), where $\rho_{XY}$ follows a harmonic behavior. Interestingly, the cusp-like feature and its variation with Co-doping cannot be explained by Stoner-Wolfarth model[17] or due to planar Hall effect under applied $H$. Thus, our experimental observations from $\rho_{XY}$ vs. $H_X$ and peculiar angular dependencies over the entire range of Co-doping with varying magnetic properties are a strong indication of the existence of an unconventional magnetoresistive feature when $H$ is applied perpendicular to the magnetic easy-axis of the vdW FM.

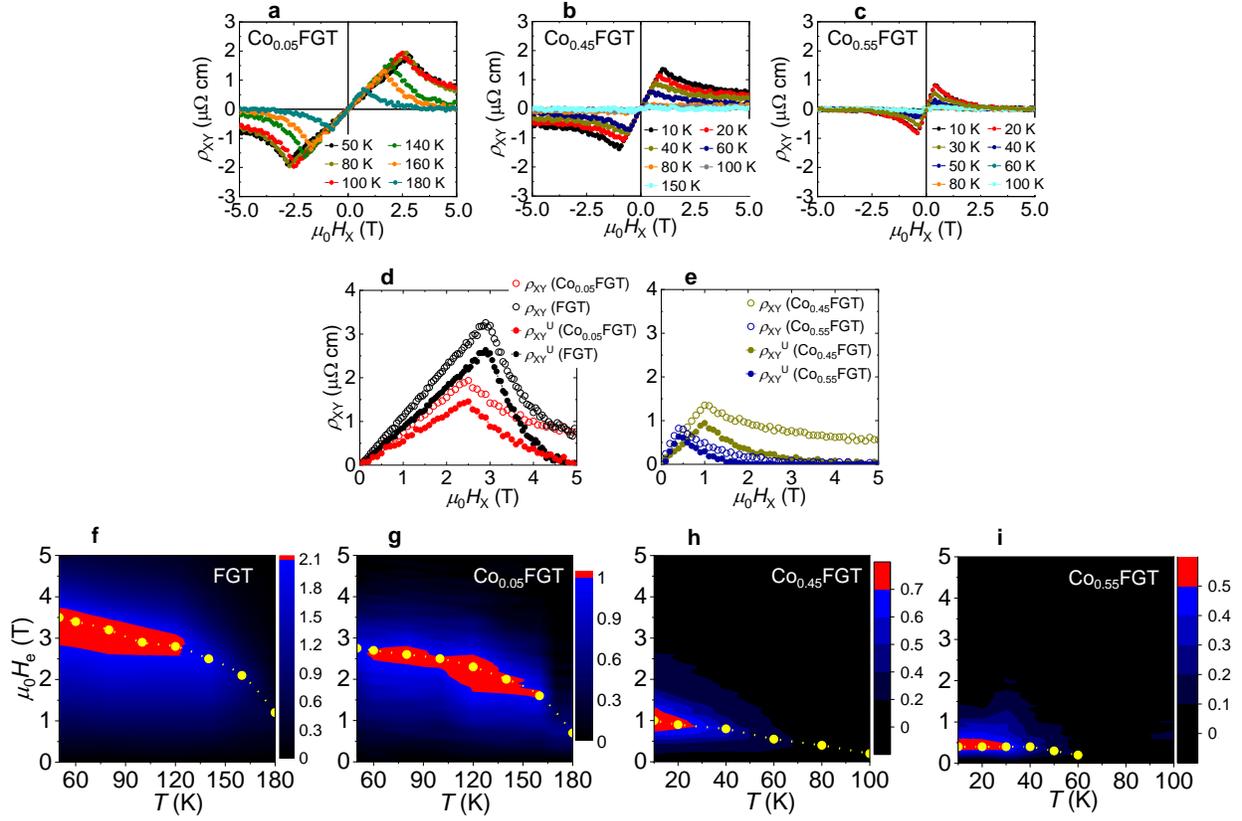

**Fig. 5.** (a), (b), (c) Hall resistivity ($\rho_{XY}$) versus applied $H_X$ (⊥ c-axis || $I$) for Co$_{0.05}$FGT, Co$_{0.45}$FGT, and Co$_{0.55}$FGT single-crystal, respectively, at various temperatures. (d) Experimental data for $\rho_{XY}$ for Co$_{0.05}$FGT (red-rimmed circle) and FGT (black-rimmed circle) at 100K, and extracted unconventional Hall contribution ($\rho_{XY}^U$) to $\rho_{XY}$ for Co-FGT (solid red circle) and FGT (solid black circle), respectively at 100 K. (e) Experimental results of similar measurements for Co$_{0.45}$FGT (dark yellow-rimmed circle) and Co$_{0.55}$FGT (blue-rimmed circle) at 10 K. (f), (g), (h), (i) Contour mapping of extracted $\rho_{XY}^U$ as a function of the applied magnetic field ($H$) and temperature ($T$) for reference FGT, Co$_{0.05}$FGT, Co$_{0.45}$FGT, and Co$_{0.55}$FGT, respectively. Yellow lines in (f)-(i) correspond to the internal emergent magnetic field ($H_e$) for the single-crystals at various temperatures.

To examine the temperature-magnetic field phase space evolution of the unconventional Hall effect and to establish a clear picture, we have examined the temperature dependence of $\rho_{XY}$ vs. $H_X$. Figure 5a-c shows the results for $Co_{0.05}FGT$, $Co_{0.45}FGT$, and $Co_{0.55}FGT$, respectively. For all the single-crystals, the cusp-like behavior persists over the entire temperature range (up to $T_C$), indicating a magnetic origin of the observed features. An increase of $x$ has a two-fold response; monotonic reduction of $\rho_{XY}$ amplitude and reduction of the magnetic field strength corresponding to maximum $\rho_{XY}$, possibly reflecting the changes in the underlying interactions as a function of doping. To extract the unconventional contribution from total Hall resistivity ($\rho_{XY}$), we consider

$$\rho_{XY} = \mu_0 R_0 H + S_A \rho_{XX}^2 M + \rho_{XY}^U, \tag{1}$$

where $\rho_{XX}$ is the longitudinal resistivity of the single crystal, $S_A$ is the field independent coefficient to the anomalous Hall resistivity determined from fitting (see supplementary figure S2), and $R_0$ is the ordinary Hall coefficient, determined from the slope of $\rho_{XY}$ vs. $H_Z$ measurements. In this convention, the first term of Eq. (1) corresponds to the ordinary Hall effect (OHE), the second term corresponds to AHE, and $\rho_{XY}^U$ corresponds to the unconventional Hall effect comprising THE and magnetoresistive contributions from topologically trivial and non-trivial underlying spin configurations. Figure 5d, e shows the extracted unconventional Hall resistivity ($\rho_{XY}^U$) from $\rho_{XY}$ and in-plane magnetization curves (see Fig. 3c, e, g, i) for $Co_{0.05}FGT$, FGT at 100 K and $Co_{0.45}FGT$, $Co_{0.55}FGT$ at 10 K, respectively. These experimental results were then utilized to constitute the phase diagram of $\rho_{XY}^U$ over a wide temperature range for all the single crystals (see Fig. 5f-i). For $Co_{0.05}FGT$ and reference FGT, a significant strength of $\rho_{XY}^U$ is obtained, larger than THE strength either in vdW material[9,18] or other potential skyrmionic FMs[19,26-32,48]. An increase in $x$ results in a reduction of $\rho_{XY}^U$ over the entire temperature range, indicating its sensitivity to the underlying interactions. The magnitude of the in-plane field at which $\rho_{XY}^U$ attains a maximum is considered to be a measure of a possible internal emergent magnetic field ($H_e$), and its strength monotonically decreases with increasing $x$. These results of an unconventional Hall effect and the ability to control the internal emergent magnetic field by doping deepens our understanding of magnetoresistive effects in 2D vdW FMs.

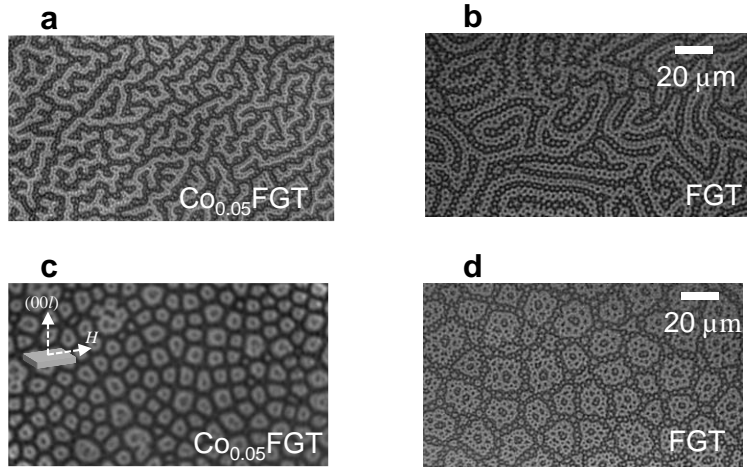

**Fig. 6.** (a), (b) Differential p-MOKE image of remnant magnetic state for $Co_{0.05}$FGT and FGT, respectively at 100 K. (c), (d) Differential p-MOKE image of remnant magnetic state for $Co_{0.05}$FGT and FGT, respectively at 100 K, obtained after field cooling under applied $\mu_0 H$ = 20 mT ($H \perp$ c-axis). The inset in (a) denotes the direction of the cooling field $H$ relative to the c-axis of the crystal. Black (grey) areas in (a)-(d) correspond to magnetization pointing in parallel (antiparallel) direction with respect to c-axis (out of the plane of paper).

**Magneto-optical imaging of domain structures.** To clarify the contribution of the underlying magnetic configuration towards the observed magnetotransport behavior, temperature-dependent MOKE measurements were carried out on the single crystals. First, a reference image was taken at 250K, well above $T_C$ of the single crystals. Then, the samples were cooled down to low temperatures (~ 50 K) in the absence of an applied magnetic field. Subsequently, images were acquired at several temperatures in the warming cycle. To enhance the signal-to-noise ratio, the difference between these MOKE images and the reference image was obtained. Note that no domain patterns were observed for $Co_{0.45}FGT$ and $Co_{0.55}FGT$ samples, indicating that the domain period is smaller than the resolution of our MOKE microscope (~ 1 µm). Thus, we restrict the experimental results of this section to that for $Co_{0.05}FGT$ and FGT single-crystals. Figure 6a, b shows the zero-field cooled (ZFC) MOKE images for $Co_{0.05}FGT$ and reference FGT at 100 K. For both the single-crystals, below $T_C$, a spontaneous generation of stripe-like surface domain patterns was observed with alternating contrast corresponding to the magnetization being parallel or antiparallel to the out-of-plane direction (*i.e.*, (00*l*) axis). Rows of circular domains or skyrmion bubble-like features with opposite magnetization orientation are embedded within this stripe domain background, as has been observed for FGT using other techniques, as well[8,15,43]. On the other hand, a remarkable transformation of the ZFC domain pattern was observed under a magnetic field $H$ ($\perp$ *c*-axis) applied perpendicular to the easy axis. For this measurement, after acquiring a reference image at 250 K, the samples were cooled to 50 K in the presence of an applied magnetic field $\mu_0 H$ ($\perp$ *c*-axis) = 20 mT (see inset of Fig. 6c) (maximum applicable magnetic field of our set-up). Subsequently, the field was turned off and MOKE images were acquired in the remnant state at several temperatures. Figures 6c, d shows the zero-field MOKE images for $Co_{0.05}FGT$ and FGT, respectively, obtained in the warming cycle after field cooling. As opposed to the stripe-like domain background in the ZFC condition, the observed domain patterns for this case can be considered as a combination of magnetic domain walls connected in rings consisting of a collection of dense circular domains or skyrmion-like bubbles[15,16]. Note that the size of the bubble-like features in our single-crystalline samples are much larger than the hexagonal skyrmion lattices observed in other studies[15,16]. Once these unconventional spin structures are generated by $H$ ($\perp$ *c*-axis), the structure remains unchanged over the entire temperature range. An increase of temperature, up to $T_C$, only results in a weakening of the magnetic contrast due to decreased magnetization and enhanced thermal effects. Besides, these textures in our $Co_{0.05}FGT$ and FGT vdW materials show remarkable stability up to much higher temperatures (~ $T_C$), unlike previous studies, which reported the existence of skyrmion-lattice phases confined within small ranges of temperature and magnetic field for chiral MnSi[29,39], $Cu_2OSeO_3$[28], and $(Ca,Ce)MnO_3$[39,48] magnets or $Gd_2PdSi_3$[30] centrosymmetric frustrated magnets.

**Discussion**

Here, the possible scenarios contributing to the observation of an unconventional Hall effect and its correlation to the complicated domain configurations from MOKE are discussed. As stated before, the occurrence of a cusp-like feature in $\rho_{XY}$ for applied $H$ deviating from the (00$l$) direction is an unresolved issue attributed either to THE from the emergent magnetic field of underlying spin textures[9,18], or to the effect of multi-domain to single-domain transformations under perpendicular magnetic anisotropy (PMA)[17]. The former is supported by the observation of sub-micrometer scale skyrmion lattice structure[15,16], while, for the latter, the experimental results from FGT were compared with similar experiments and micromagnetic simulations on Ta/CoFeB/MgO multilayers[17]. The present experimental results with Co-doping reveals subtle differences, which enables the identification of the dominant contribution towards this unconventional magnetoresistive feature. Our temperature-dependent magnetization demonstrates a monotonic reduction of $T_C$ ($T_C$ ~ 220 K for FGT to ~ 50 K for $Co_{0.55}$FGT), and a significant reduction of PMA with increasing $x$. Under the scenario of multi-to-single domain transformation[17], a reduction of PMA is only expected to result in decreased $\rho_{XY}$ amplitude under applied $H$ ($\perp$ $c$-axis), while $H_e$ remains virtually constant. This is expected since the reduction of PMA is associated with a stronger tendency to align the magnetization along the in-plane directions under applied $H$ ($\perp$ $c$-axis). On the other hand, our experimental results, along with the reduction of $\rho_{XY}$ amplitude also show a significant reduction of $H_e$ (from 3.5 T in FGT to ~ 0.3 T in $Co_{0.55}$FGT). Complementary MOKE measurements also reveal a complicated spontaneous magnetic domain pattern consisting of bubble-like features which was converted into an unconventional ring-like structure with applied $H$ perpendicular to the (00$l$) easy directions. We attribute these ring-like structures to an aggregate of connected topological skyrmion bubble-like or lattice structures and magnetic domain walls along with possible existence of trivial circular and/or stripe domains. Under this scenario, assuming that the ring-like patterns are composed of sub-micrometer sized bubble-like structures, the position of the cusp-like feature in our magnetotransport measurements corresponds to the emergent field from the underlying spin textures, attributed to the total flux quantum contained in these structures. The substitution of Fe by Co results in a modification of magnetic exchange and anisotropy, along with the introduction of symmetry-breaking interactions in the bulk[44], leading to a modification of the spin configurations and associated flux quanta manifesting in significant depreciation of the internal emergent magnetic field. Future experimental investigations on imaging of the magnetic configurations at higher $H$ and/or theoretical investigations on magnetic textures in vdW FMs would provide significant insights required for further understanding of the complicated spin structures in vdW FMs. Our results offer a route towards the realization of new-concept spin textures in vdW FMs, promising for non-collinear spin texture based physics and spintronic devices.

**Method**

**Single-crystal growth.** The single crystals of Co$_x$FGT and reference FGT were grown by CVT method with $I_2$ as a transport agent. Starting from a mixture of pure elements Fe (5N), Ge (5N), Te (5N) and Co (5N), the mixture was sealed in an evacuated quartz tube and heated in a two-zone furnace with a temperature gradient 750/700° C for one week. Large crystals of sizes up to ~ $12 \times 8$ mm$^2$ were obtained, cleavable in the *ab* plane.

**Data availability**

The data which support the findings of this work are available from the corresponding authors upon reasonable request.

**Acknowledgements**

The authors thank Dr. Takaaki Dohi for discussions. RRC acknowledges Department of Science and Technology (DST), Government of India, for financial support (Grant no. DST/INSPIRE/04/2018/001755). RPS acknowledges Science and Engineering Research Board (SERB), Government of India, for Core Research Grant CRG/2019/001028. A portion of this work was supported by JSPS Kakenhi 19H05622, 20K15155 and RIEC International Cooperative Research Projects, Tohoku University. O.A.T. acknowledges the support by the Australian Research Council (Grant No. DP200101027) and NCMAS grant.

**Author contributions**

RRC and RPS planned the study. RRC synthesized the single-crystal samples and carried out structural and magnetic characterizations. RRC and S.D. performed the electrical and magneto-optical measurements and analyzed the data with inputs from O.A.T. RRC and S.D. wrote the manuscript with input from all the authors. All authors discussed the results.

**Competing interests**

The authors declare no competing financial interests.

**Author information**

Reprints and permission information is available at www.nature.com/reprints. The authors declare no competing financial interests. Readers are welcome to comment on the online version of the paper. Correspondence and requests for materials should be addressed to RRC (rajeswari@iiserb.ac.in) or S.D. (sdg@riec.tohoku.ac.jp) or RPS (rpsingh@iiserb.ac.in).